\RequirePackage[2020-02-02]{latexrelease}
\documentclass[twocolumn,preprintnumbers]{revtex4}%
\usepackage{amsopn}
\usepackage{amssymb}
\usepackage{amsmath}
\usepackage{units}
\usepackage{graphicx}
\usepackage{dcolumn}
\usepackage{bm}
\usepackage{xcolor}
\usepackage{ifthen}
\usepackage[normalem]{ulem}
\usepackage{amsfonts}%
\RequirePackage[2020-02-02]{latexrelease}
\UseRawInputEncoding
\providecommand{\U}[1]{\protect\rule{.1in}{.1in}}
\providecommand{\U}[1]{\protect\rule{.1in}{.1in}}
\def\showal{1}
\newcommand{\al}[1]{\ifthenelse{\showal=1}{\textcolor{orange}{[[#1]]}}{}}

\newcommand{\eb}[1]{\ifthenelse{\showal=1}{\textcolor{cyan}{[[#1]]}}{}}

\begin{document}

\title{Tunable multimode lasing in a fiber ring}
\author{Eyal Buks}
\affiliation{Andrew and Erna Viterbi Department of Electrical Engineering, Technion,
Haifa 32000 Israel}
\date{\today }

\begin{abstract}
We experimentally study a fiber loop laser with an integrated
Erbium doped fiber (EDF). The output optical spectrum is
measured as a function of the EDF temperature. We find that
below a critical temperature of about 10K the measured optical spectrum
exhibits a sequence of narrow and unequally-spaced peaks. Externally
injected light and filtering are employed for tuning the peaks' wavelengths. Operation of the device as an optical memory having storage time of about $\unit[20]{ms}$ is demonstrated. The multimode lasing tunability can be exploited for novel applications in
the fields of sensing, communication, and quantum data storage.
\end{abstract}

\maketitle

\textbf{Introduction} - Erbium doped fibers (EDF) are widely employed in a
variety of applications. Key properties of EDF can be controlled by varying
the temperature \cite{Kobyakov_1,Le_3611,Thevenaz_22}. Multimode lasing has
a variety of applications in the fields of sensing, spectroscopy, signal
processing and communication \cite{Haken_laser}. Multimode lasing in the
telecom band has been demonstrated by integrating an EDF cooled by liquid
nitrogen into a fiber ring laser \cite%
{Yamashita_1298,Liu_102988,Lopez_085401}. It has been recently proposed that
EDF operating at low temperatures can be used for storing quantum
information \cite%
{Saglamyurek_83,Staudt_720,Wei_2209_00802,Ortu_035024,Liu_2201_03692,Veissier_195138}%
. Population storage times exceeding 10 seconds \cite{Saglamyurek_241111},
and electronic spin lifetimes exceeding one hour \cite{Shafiei_F2A}, have
been demonstrated using persistent spectral hole burning.

\begin{figure}[tp]
\begin{center}
\includegraphics[width=3.2in,keepaspectratio]{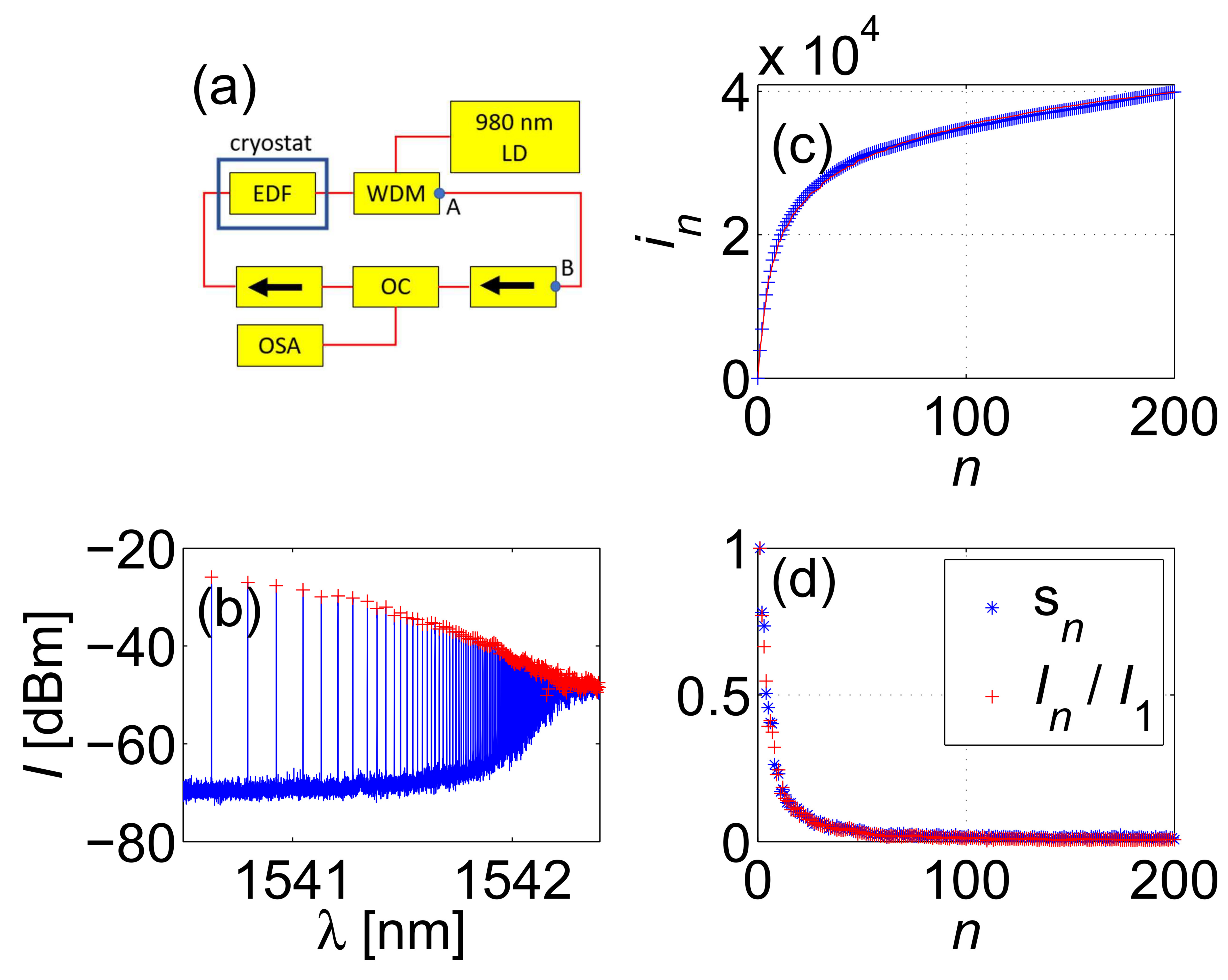}
\end{center}
\par
.
\caption{{}USOC - (a) The experimental setup. (b) The measured optical
spectrum with diode current $I_{\mathrm{D}}=\unit[120]{mA}$. (c) Comparison
between the measured normalized frequencies $i_{n}=\left( f_{0}-f_{n}\right)
/f_{\mathrm{L}}$ and the calculated values of $\protect\nu \log p_{n}$ [see
Eq. \protect\ref{i_k}]. The dimensionless pre-factor $\protect\nu $ is found
by fitting to be given by $\protect\nu =5610$. (d) Comparison between the
normalized intensities $I_{n}/I_{1}$ and the normalized wavelength gaps $%
\left( \protect\lambda _{n+1}-\protect\lambda _{n}\right) /\left( \protect%
\lambda _{1}-\protect\lambda _{0}\right) $ [see Eq. \protect\ref{s_k}].}
\label{FigComb}
\end{figure}

In this work we study a fiber loop laser with an integrated EDF \cite%
{Antuzevics_1149,haken1985laser}. We measure the emitted optical spectrum as
a function of the EDF temperature \cite{Aubry_2100002,Pizzaia_2352}. Below a
critical temperature of about $\unit[10]{K}$ the measured optical spectrum
exhibits an unequally-spaced optical comb (USOC) made of a sequence of
narrow peaks. The observed USOC and multimode lasing are attributed to
intermode coupling \cite{moloney2018nonlinear,Buks_128591}. We find that the
wavelengths at which the multimode lasing occurs can be controlled by light
that is externally injected into the fiber loop. The possibility of using the device under study as an optical memory is explored, and a storage time of about $\unit[20]{ms}$ is demonstrated. The USOC controllability,
together with its high stability, can be exploited for facilitating novel
applications in the fields of sensing, communication, and quantum data storage.

\textbf{Experimental setup} - The experimental setup is schematically
depicted in Fig. \ref{FigComb}(a). EDF having length of $\unit[20]{m}$,
absorption of $30$ dB $\unit{m}^{-1}$ at $\unit[1530]{nm}$, and mode field
diameter of $6.5\mu \unit{m}$ at $\unit[1550]{nm}$, is cooled down using a
cryogen free cryostat. The EDF is thermally coupled to a calibrated silicon
diode serving as a thermometer, and it is pumped using a $\unit[980]{nm}$
laser diode (LD) biased with current denoted by $I_{\mathrm{D}}$. The cold
EDF is integrated with a room temperature fiber loop using a
wavelength-division multiplexing (WDM) device. Two isolators [labeled by
arrows in the sketch shown in Fig. \ref{FigComb}(a)] and a 10:90 output
coupler (OC) are integrated in the fiber loop. The loop frequency $f_{%
\mathrm{L}}$ (inverse loop period time), which is measured using a radio
frequency spectrum analyzer and a photodetector, is given by $f_{\mathrm{L}%
}=c/\left( n_{\mathrm{F}}l_{\mathrm{L}}\right) =\unit[4.963]{MHz}$, where $c$
is the speed of light in vacuum, $n_{\mathrm{F}}=1.45$ is the fiber
refractive index, and $l_{\mathrm{L}}=\unit[41.7]{m}$ is
the fiber loop total length. An optical spectrum analyzer (OSA) is connected
to the 10:90 OC.

\textbf{USOC} - Near wavelength of $\unit[1540]{nm}$ and below a critical
temperature of about $\unit[10]{K}$ the measured optical spectrum exhibits
narrow peaks at a sequence of wavelengths denoted by $\left\{ \lambda
_{n}\right\} $, where $n=0,1,2,\cdots $ [see Fig. \ref{FigComb}(b)].
For the data presented in Fig. \ref{FigComb} $\lambda _{0}=\unit[1540.629]{nm}$
and $\lambda _{1}-\lambda _{0}=\unit[0.1661]{nm}$. The frequency $f_{n}$
associated with wavelength $\lambda _{n}$ is given by $f_{n}=c/\lambda _{n}$%
. The intensity of the USOC peak occurring at wavelength $\lambda _{n}$ is
denoted by $I_{n}$.

\begin{figure}[tbp]
\begin{center}
\includegraphics[width=3.3in,keepaspectratio]{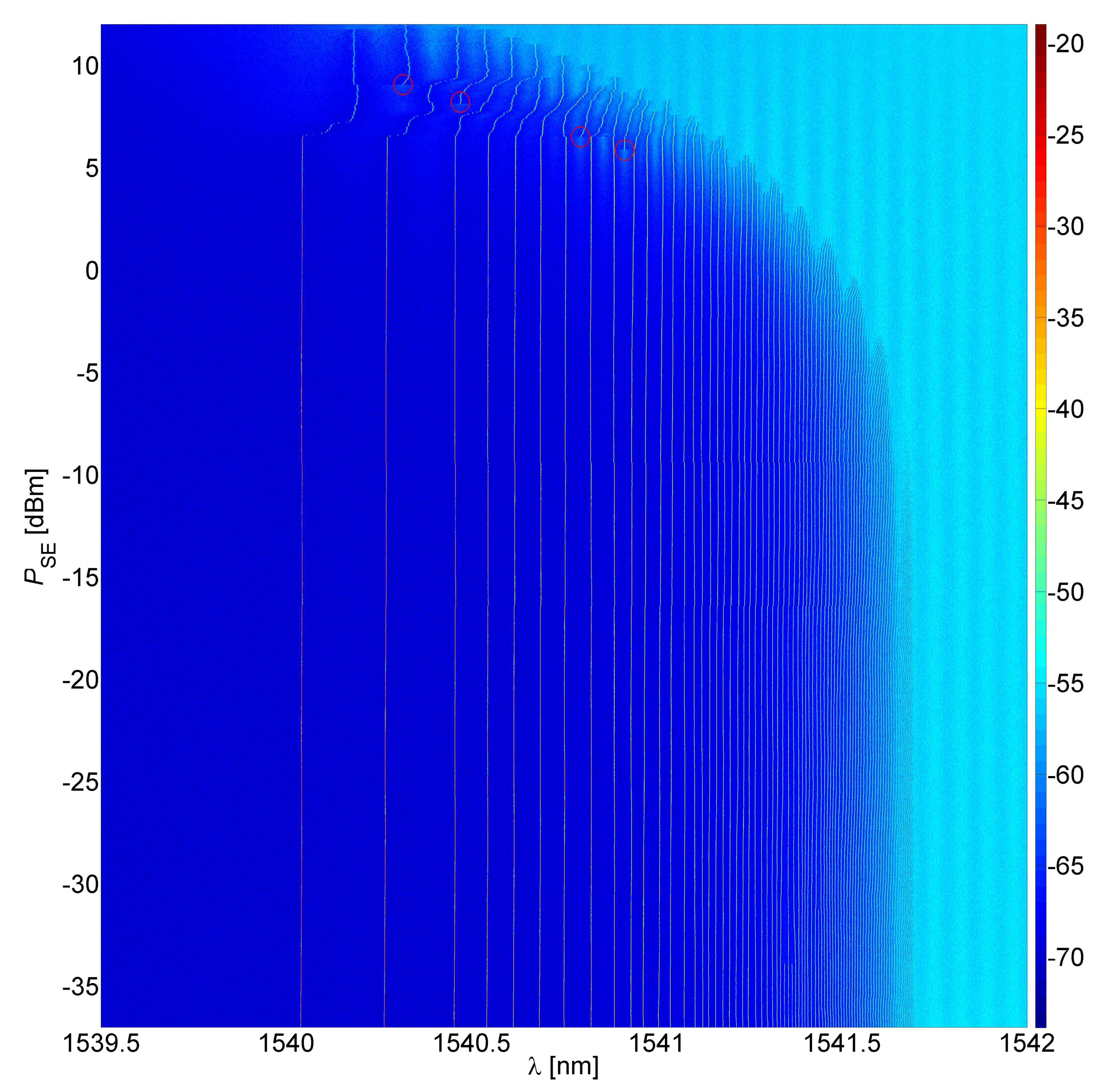} 
\end{center}
\caption{{}Gain saturation. Optical spectrum (in dBm units) as a function of
the SE optical power $P_{\mathrm{SE}}$ feeding the 1:99 OC. The temperature
is $T=\unit[2.9]{K}$ and diode current is $I_{\mathrm{D}}=\unit[100]{mA}$.}
\label{FigGainSat}
\end{figure}

\begin{figure}[tbp]
\begin{center}
\includegraphics[width=2.8in,keepaspectratio]{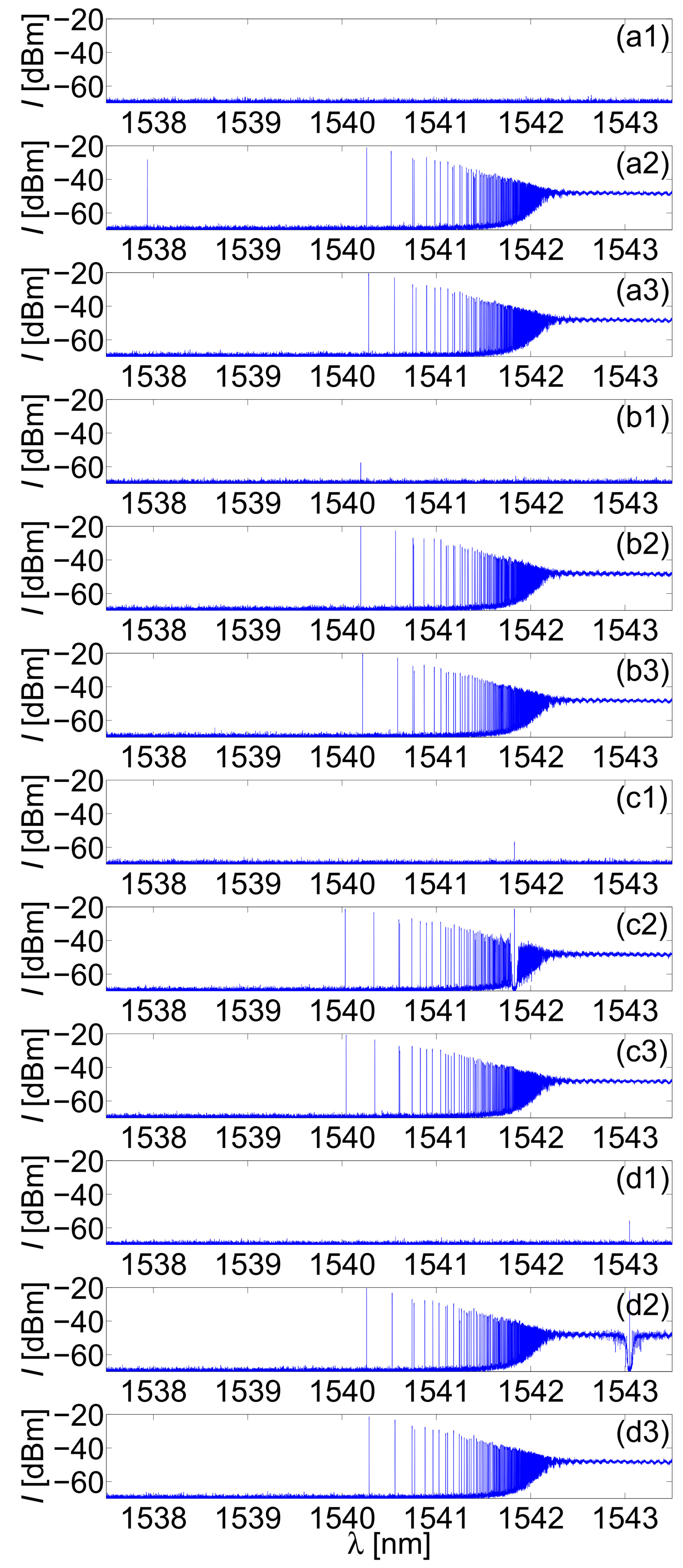} 
\end{center}
\caption{{}Laser injection. The measured optical spectrum after the first
(1), second (2) and third (3) step. The monochromatic injected light of the
external laser has wavelength $\protect\lambda _{\mathrm{L}}$ of (a) $%
\protect\lambda _{\mathrm{L}}=\unit[1537.917]{nm}$, (b) $\protect\lambda _{ 
\mathrm{L}}=\unit[1540.162]{nm}$, (c) $\protect\lambda _{\mathrm{L}}= 
\unit[1541.784]{nm}$ and (d) $\protect\lambda _{\mathrm{L}}=\unit[1543.000]{nm}$. The
temperature is $T=\unit[3.0]{K}$, diode current is $I_{\mathrm{D}}=\unit[200]{mA}
$, and tunable laser power is $\unit[0.1]{mW}$.}
\label{FigCombW}
\end{figure}

\begin{figure}[tbp]
\begin{center}
\includegraphics[width=3.3in,keepaspectratio]{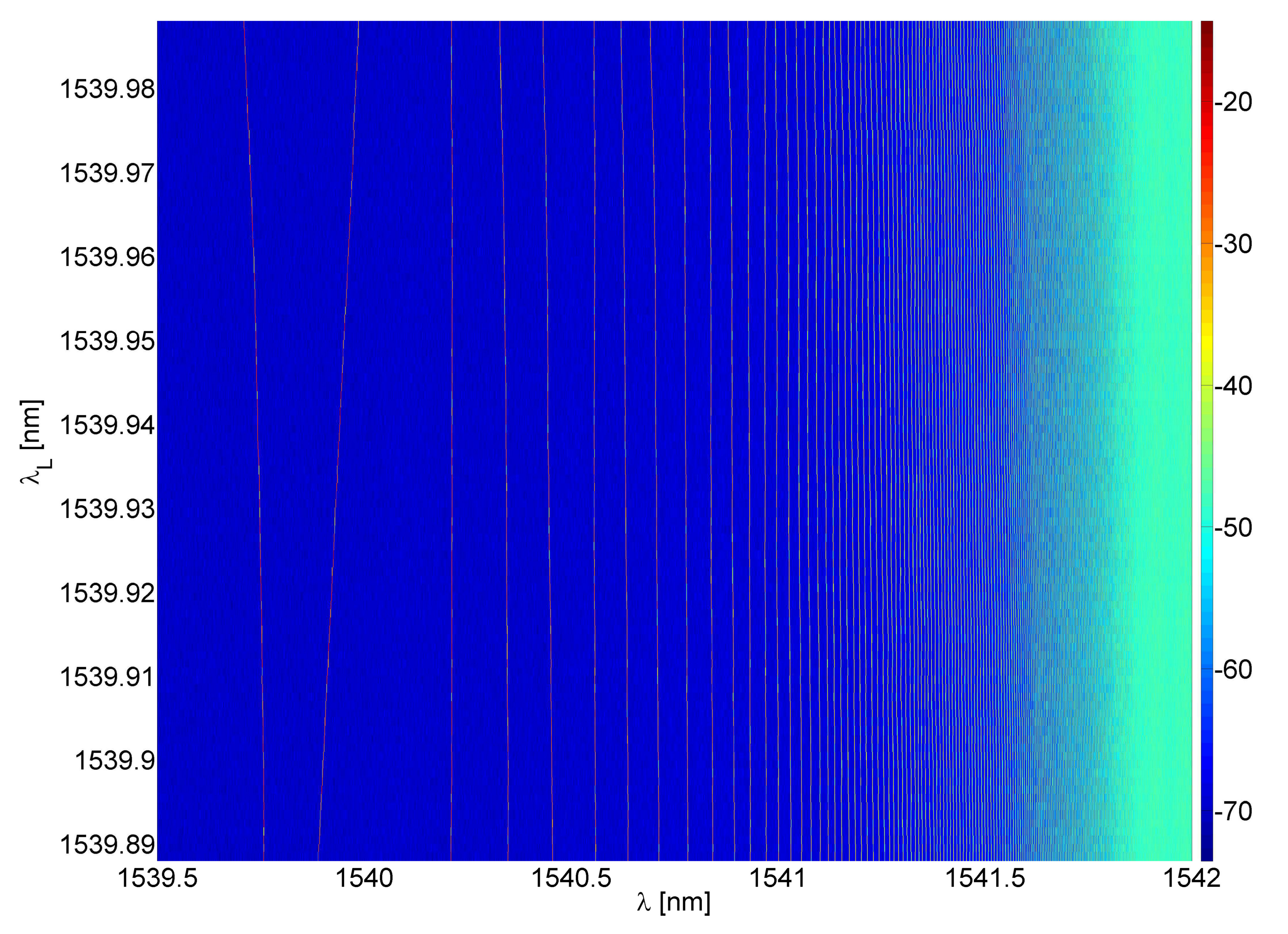} 
\end{center}
\caption{Peak moving. The tunable laser wavelength $\protect\lambda _{ 
\mathrm{L}}$ is varied from $\unit[1539.988]{nm}$ to $\unit[1539.888]{nm}$ in
steps of $\unit[1]{pm}$. During the time when the wavelength $\protect\lambda %
_{\mathrm{L}}$ is modified, the external laser optical power is turned off
(using an external attenuator). The measured optical spectrum is shown in
dBm units as a function of $\protect\lambda _{\mathrm{L}}$. The temperature
is $T=\unit[2.9]{K}$, diode current is $I_{ \mathrm{D}}=\unit[200]{mA}$, and
tunable laser power is $\unit[0.39]{mW}$.}
\label{FigCombM}
\end{figure}

\begin{figure}[tbp]
\begin{center}
\includegraphics[width=3.2in,keepaspectratio]{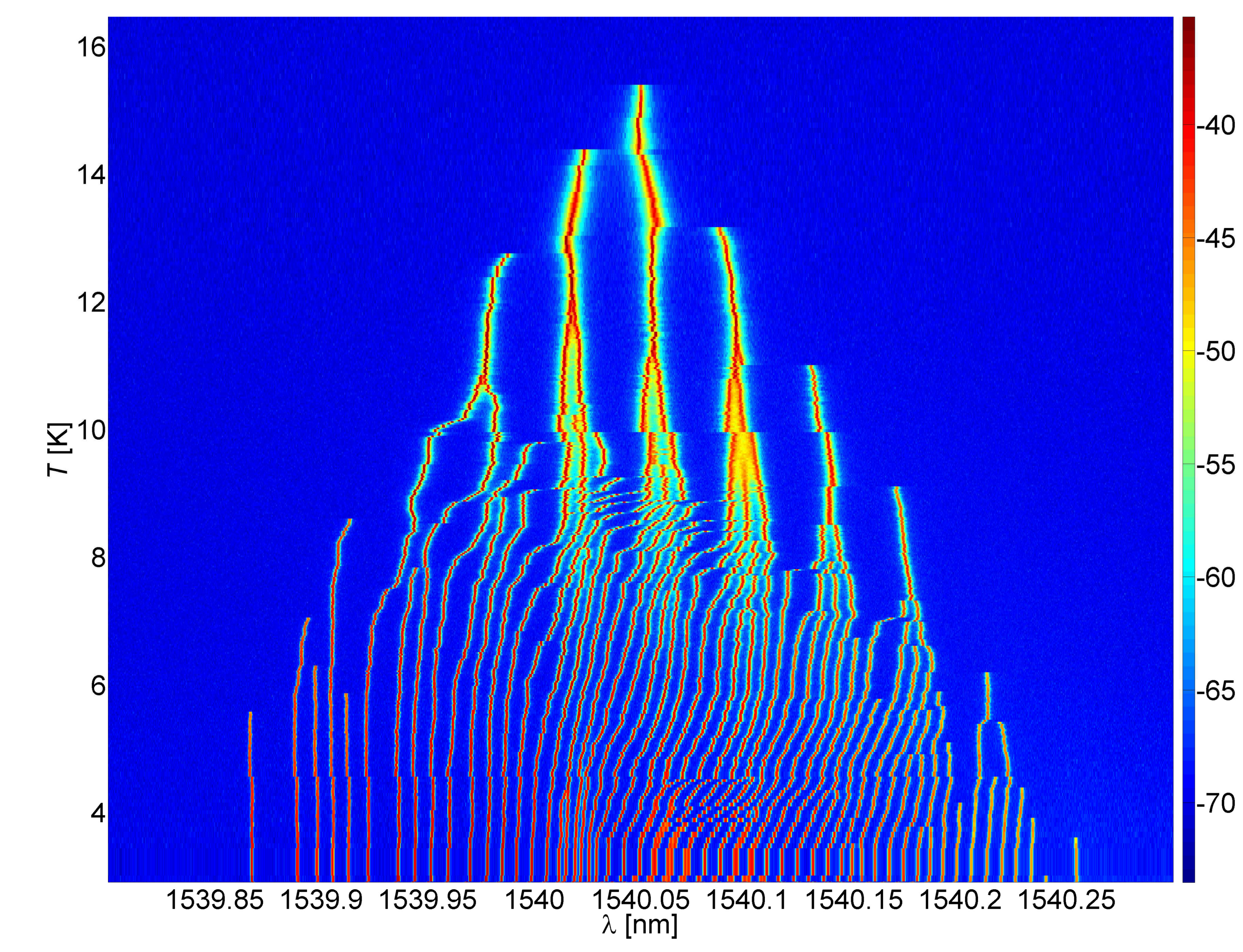} 
\end{center}
\caption{{}Optical filter. A tunable filter is integrated into the fiber
loop. The plot displays the optical spectrum (in dBm units) as a function of
temperature $T$. The optical filter central wavelength $\protect\lambda _{ 
\mathrm{F}}$ is tuned to the value $\protect\lambda _{\mathrm{F}}=\unit[1540]{
nm}$, and the diode current is set to $I_{\mathrm{D}}=\unit[100]{mA}$.}
\label{FigTF}
\end{figure}

\begin{figure}[tbp]
	\begin{center}
		\includegraphics[width=3.2in,keepaspectratio]{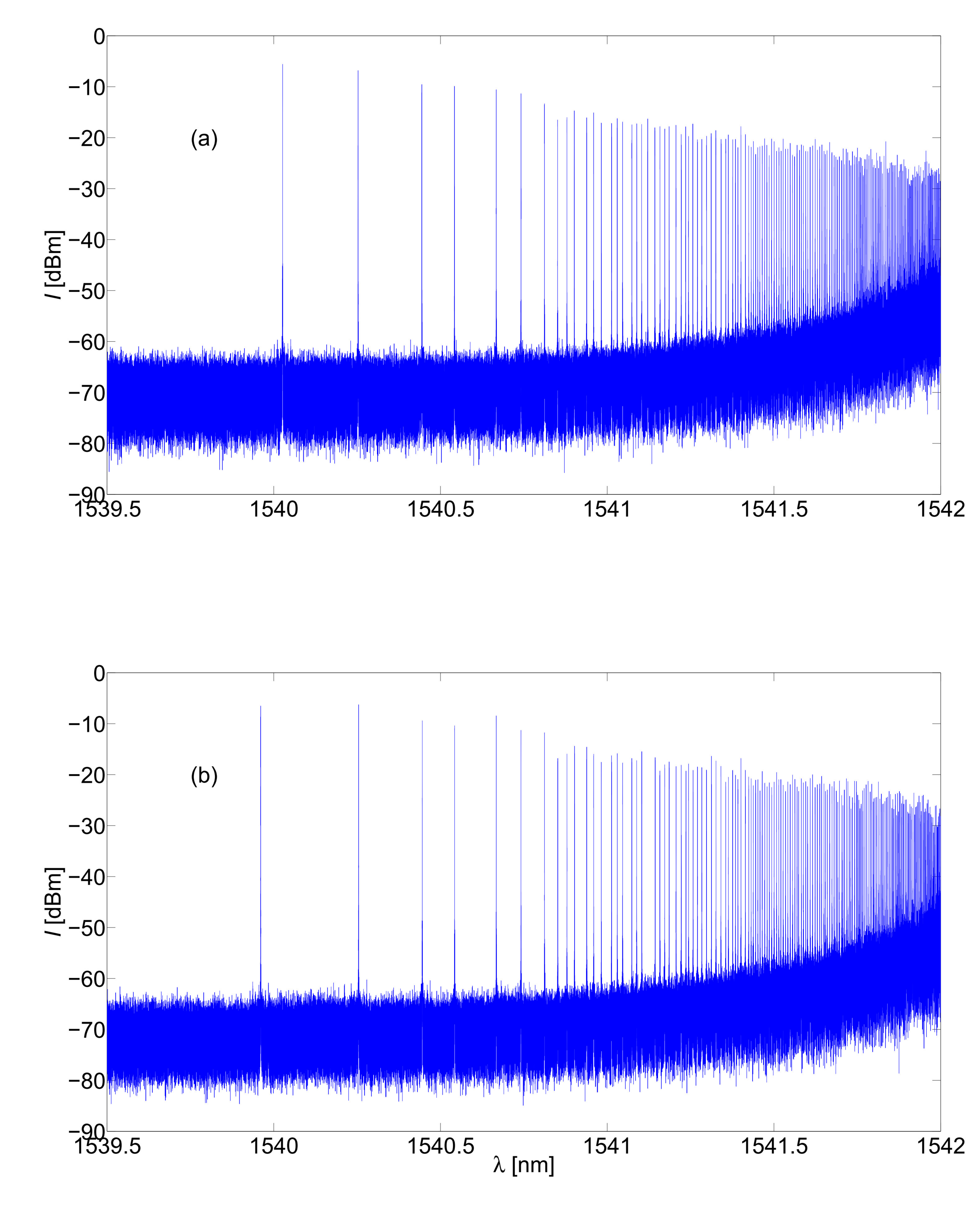} 
	\end{center}
	\caption{{}USOC memory. Optical spectrum before (a) and after (b) an off pulse having time duration of $\tau_{\mathrm{off}}=\unit[22]{ms}$. In our experimental setup, the LD bias current $I_{\mathrm{D}}$
		can be fully modulated with frequency up to of about $ \unit[150]{kHz}$. Diode current (before and after the off pulse) is $I_{\mathrm{D}}=\unit[150]{mA}
		$, and temperature is $T=\unit[3.5]{K}$. }
	\label{FigMem}
\end{figure}

The normalized frequency sequence $i_{n}\equiv \left( f_{0}-f_{n}\right) /f_{%
\mathrm{L}}$, and the normalized intensity sequence $s_{n}\equiv I_{n}/I_{1}$
are well describe by the following empirical laws%
\begin{eqnarray}
i_{n} &=&\nu \log p_{n}\ ,  \label{i_k} \\
s_{n} &=&\frac{\lambda _{n+1}-\lambda _{n}}{\lambda _{1}-\lambda _{0}}\ ,
\label{s_k}
\end{eqnarray}%
where $\nu $ is a positive constant, and $p_{n}$ is the $n$'th prime number.
The comparison between the measured values of $i_{n}=\left(
f_{0}-f_{n}\right) /f_{\mathrm{L}}$ and the calculated values of $\nu \log
p_{n}$ [see Eq. (\ref{i_k})] yields a good agreement [see Fig. \ref{FigComb}%
(c)]. The level of agreement is quantified by the parameter $\varepsilon =n_{%
\mathrm{m}}^{-1}\sum_{n=1}^{n_{\mathrm{m}}}\left\vert \left( i_{n}-\nu \log
p_{n}\right) /i_{n_{\mathrm{m}}}\right\vert $, where $n_{\mathrm{m}}$ is the
number of peaks that can be reliably resolved. For the data shown in Fig. (%
\ref{FigComb}) $n_{\mathrm{m}}=200$ and $\varepsilon =0.0043$. The
comparison between $I_{n}/I_{1}$and $\left( \lambda _{n+1}-\lambda
_{n}\right) /\left( \lambda _{1}-\lambda _{0}\right) $ [see Eq. (\ref{s_k})]
is shown in Fig. \ref{FigComb}(d). The underlying mechanism responsible for USOC formation has remained mainly
unknown. The connection between the normalized frequency sequence $%
i_{n}=\left( f_{0}-f_{n}\right) /f_{\mathrm{L}}$ and the sequence of prime
numbers [see Eq. (\ref{i_k})] is discussed in \cite{Buks_128591}.

\textbf{Gain saturation} - The fiber loop gain can be characterized by
injecting input light having a continuous spectrum. Injection measurements
[shown below in Figs. \ref{FigGainSat}, \ref{FigCombW} and \ref{FigCombM}]
are performed by integrating a 1:99 OC into the loop between the points
labeled as 'A' and 'B' in Fig. \ref{FigComb}(a). For the measurements
presented in Fig. \ref{FigGainSat} below, the 1:99 OC is employed to inject
into the loop spontaneous emission (SE) light from an external pumped EDF.
The optical power of the externally injected SE light is controlled by a
tunable attenuator. The plot shown in Fig. \ref{FigGainSat} exhibits the
spectrum measured by the OSA as a function of the SE optical power $P_{%
\mathrm{SE}}$ feeding the 1:99 OC.

Consider two neighboring USOC peaks at wavelength $\lambda _{n}$ and $%
\lambda _{n+1}$, where $n$ is a non-negative integer. As can be seen from
Fig. \ref{FigGainSat}, the loop gain $g_{\mathrm{L}}$ in the region $\left[
\lambda _{n},\lambda _{n+1}\right] $ is approximately a symmetric function
around the mid point $\lambda _{\mathrm{a},n}=\left( \lambda _{n}+\lambda
_{n+1}\right) /2$, at which $g_{\mathrm{L}}$ peaks. The loop gain $g_{%
\mathrm{L}}$ vanishes at the end points $\lambda _{n}$ and $\lambda _{n+1}$.

When the SE input optical power $P_{\mathrm{SE}}$ is sufficiently high, new USOC peaks at the
mid point wavelengths $\lambda _{\mathrm{a},n}$ are created (see the regions
highlighted by the overlaid red circles in Fig. \ref{FigGainSat}). This
behavior is attributed to the fact that the wavelengths $\lambda _{n}$ and $%
\lambda _{n+1}$, together with the mid wavelength $\lambda _{\mathrm{a},n}$,
nearly satisfy the mixing condition $f _{n}+f _{n+1}=2f_{\mathrm{a},n}$, where $f_{\mathrm{a},n}=c/\lambda _{\mathrm{a},n}$ (recall that $\left\vert \lambda _{n+1}-\lambda
_{n}\right\vert \ll \lambda _{n}$). Note that the USOC created with
relatively intense SE injection (see Fig. \ref{FigGainSat}) is significantly
distorted, and it cannot be well described by the empirical laws (\ref{i_k})
and (\ref{s_k}).

\textbf{Laser injection} - The plot shown in Fig. \ref{FigCombW}
demonstrates external tuning of an USOC peak. For the measurements presented
in Fig. \ref{FigCombW}, the 1:99 OC is fed with a narrow band external laser
having a tunable wavelength $\lambda _{\mathrm{L}}$. The experimental
protocol has three steps, In the first one the external laser having
wavelength $\lambda _{\mathrm{L}}$ is turned on, then in the second step the
diode current is turned on, and in the last step the external laser is
turned off. After each step the optical spectrum is measured. The measured
spectrum after the first, second and third step is shown in the plots in
Fig. \ref{FigCombW} labelled by the numbers 1, 2 and 3, respectively.

In some cases, the spectrum that is measured after the third step, i.e.
after turning off the external laser, contains a peak at the injected
external laser wavelength $\lambda _{\mathrm{L}}$. This is demonstrated by
plots in Fig. \ref{FigCombW} labelled by the letters b and c. We find that
this persistent effect occurs only when $\lambda _{\mathrm{L}}$ is tuned
inside the USOC region of about $\unit[1540-1542.5]{nm}$ (compare to the plots
in Fig. \ref{FigCombW} labelled by the letters a and d, for which the
persistent effect does not occur).

\textbf{Moving a single peak} - Each USOC peak wavelength $\lambda _{n}$ can
be individually tuned using external laser injection. The tuning method is
demonstrated for the second ($n=1$) USOC peak by the plot shown in Fig. \ref%
{FigCombM}. Peak moving is achieved by first tuning the external laser
wavelength $\lambda _{\mathrm{L}}$ to match the USOC peak wavelength to be
moved $\lambda _{n}$, and then by gradually tuning $\lambda _{\mathrm{L}}$
to a target value denoted by $\lambda _{n,\mathrm{T}}$. After reaching the
target value, the external laser is switched off. For the example shown in Fig. \ref{FigCombM}, all peaks with $n>1$ are unaffected by the moving process applied to the $n=1$ peak.

The moving of the USOC $n$'th peak is possible when the external laser power 
$P_{\mathrm{L}}$ exceeds a threshold value denoted by $P_{n}$. By carefully
tuning $P_{\mathrm{L}}$ (to be sufficiently close to $P_{n}$), the tuning
process of the $n$'th peak wavelength from the initial value $\lambda _{n}$
to the target value $\lambda _{n,\mathrm{T}}$ can be performed without
affecting the wavelengths $\lambda _{n^{\prime }}$ of all other USOC peaks
having $n^{\prime }\neq n$.

\textbf{Optical filter} - The band inside which USOC occurs can be
controlled using an optical filter. For the measurement shown in Fig. \ref%
{FigTF}, a Fabry--P\'{e}rot filter having full width at half maximum of $%
\unit[1.2]{nm}$ is integrated into the loop between the points 'A' and 'B' [see
the sketch in Fig. \ref{FigComb}(a)]. The plot shown in Fig. \ref{FigTF},
which exhibits the temperature dependence of the created USOC, demonstrates
a transition from the low-temperature multimode lasing, to the higher
temperature single mode lasing. Note that the same above-discussed tuning
methods (see Figs. \ref{FigCombW} and \ref{FigCombM}) can be successfully
implemented to manipulate the peaks of the filtered USOC shown in Fig. \ref{FigTF}.

\textbf{Memory} - The USOC wavelength sequence $\left\{ \lambda
_{n}\right\} $ is highly stable. Over time periods exceeding 20 hours, changes in the sequence $\left\{ \lambda
_{n}\right\} $ are too small to be reliably resolved by our OSA, provided that the $\unit[980]{nm}$ LD bias current $I_{\mathrm{D}}$ is kept constant. However, the sequence $\left\{ \lambda
_{n}\right\} $ is significantly modified each time $I_{\mathrm{D}}$ is switched off, and then switched on again, provided that the off time $\tau_{\mathrm{off}}$ is sufficiently long. On the other hand, we find that the sequence $\left\{ \lambda
_{n}\right\} $ remains unchanged when $\tau_{\mathrm{off}}$ is sufficiently short.

Amplitude modulation applied to the LD bias current $I_{\mathrm{D}}$ is employed to determine the USOC memory time, which is denoted by $\tau_{\mathrm{M}}$. This is done by measuring the sequence $\left\{ \lambda
_{n}\right\} $ before and after a switching off pulse having time duration $\tau_{\mathrm{off}}$ is applied to $I_{\mathrm{D}}$. We find that when $\tau_{\mathrm{off}} \ll \tau_{\mathrm{M}}$, the sequence $\left\{ \lambda
_{n}\right\} $ is unaffected by the off pulse, where $\tau_{\mathrm{M}}  \simeq \unit[20]{ms}$. On the other hand, $\left\{ \lambda
_{n}\right\} $ is significantly modified for $\tau_{\mathrm{off}} \gg \tau_{\mathrm{M}}$. In the region where $\tau_{\mathrm{off}}  \simeq \unit[20]{ms}$, we find that most of the wavelengths in the sequence $\left\{ \lambda
_{n}\right\} $ remain unchanged, except for the first and/or the second ones (having the shortest wavelengths). This behavior is demonstrated by the plots shown in Fig. \ref{FigMem}, for which the off time is $\tau_{\mathrm{off}}=\unit[22]{ms}$. For this case, the off pulse shifts the first wavelength $\lambda_{0}$ by $\unit[-65]{pm}$, and all other peaks are nearly unaffected. Note that the memory time $\tau_{\mathrm{M}}$ is expected to strongly depend on the temperature \cite{Saglamyurek_83,Staudt_720,Wei_2209_00802,Ortu_035024,Liu_2201_03692,Veissier_195138,Saglamyurek_241111,Shafiei_F2A}.

\textbf{Summary} - Further study is needed to reveal the underlying
mechanism responsible for the USOC formation. Our measurements demonstrate
several methods that allow tuning of the wavelengths at which multimode
lasing occurs. Future study will focus on systematically characterizing
spectral hole burning \cite{Rittner_1} in the system under study, in order
to explore its performance for quantum memory applications.

\textbf{Acknowledgments} - We thank Tian Zhong for useful discussions. This work was supported by the Israeli science foundation. 

\appendix


\bibliographystyle{ieeepes}
\bibliography{acompat,Eyal_Bib}

\end{document}